\documentstyle[prl,aps,epsf,multicol,array]{revtex}
\begin{document}
\draft
\title{Effect of the Tunneling Conductance on the Coulomb Staircase}
\author{Georg G\"oppert and Hermann Grabert}
\address{Fakult\"at f\"ur Physik, Albert--Ludwigs--Universit{\"a}t,
Hermann--Herder--Stra{\ss}e~3, D--79104 Freiburg, Germany}
\author{Nikolai V.\  Prokof'ev and Boris V.\ Svistunov}
\address{Russian Research Center ``Kurchatov Institute",
123182 Moscow, Russia}

\date{\today}
\maketitle
\widetext
\begin{abstract}
Quantum fluctuations of the charge in the single electron box are
investigated. The rounding of the Coulomb staircase caused by virtual
electron tunneling is determined by perturbation theory up to third
order in the tunneling conductance and compared with precise
Monte Carlo data computed with a new algorithm. The remarkable
agreement for large conductance indicates that presently
available experimental data on Coulomb charging effects in metallic
nanostructures can be well explained by finite order perturbative  
results.
\end{abstract}


\raggedcolumns
\begin{multicols}{2}
\narrowtext
Coulomb blockade effects in metallic nanostructures are well
understood in the region of low temperatures and for small tunneling
conductance $G_T \ll G_K=e^2/h$ \cite{nato}. Recently, several groups
have started to investigate in detail the breakdown of charging
effects. High-temperature anomalies due to weak Coulomb blockade have
been studied
\cite{pekola94,diplom,golubevJETP,goeppertPRB,joyezPRB,wangpekola,scha}, and
progress in fabrication techniques has lead to reliable
experimental data for systems with a tunneling conductance of several
$G_K$ \cite{wangpekola,scha,pekola96,joyez}. Most of the  
theoretical work on Coulomb
blockade for strong tunneling can roughly be divided into two
groups. On the one
hand, a significant body of work restricts the theory to usually two
charge states \cite{matveev,falci,schoeller}. This
requires the introduction of an arbitrary cut-off that enters the final
results and deranges the comparison with experimental findings. On the
other hand, systematic perturbative calculations in powers of the
tunneling Hamiltonian $H_T$ can be shown
\cite{physrevB} to be independent of the electronic bandwidth and give
results in terms of experimentally measurable quantities. While this
latter approach was successful \cite{koenig,koenigpreprint} in  
explaining some of the
recent data on strong tunneling in the single electron transistor
\cite{joyez}, the range of validity of perturbative
results is not known a priori. In this Letter, we focus on the simplest system
displaying charging
effects, the single electron box. Perturbative results up to
sixth order in $H_T$ are derived and compared with Monte Carlo
data. We shall demonstrate that perturbation theory indeed does
remarkably well.
\begin{figure}
\begin{center}
\leavevmode
\epsfxsize=0.25\textwidth
\epsfbox{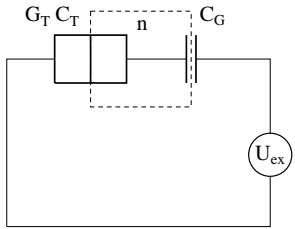}
\end{center}
\caption{Circuit diagram of the single electron box, consisting of
a tunnel junction in series with a capacitor.}
\label{fig:fig1}
\end{figure}
The circuit diagram of the single electron box in
Fig.~\ref{fig:fig1} shows a tunnel junction with capacitance $C_T$
and tunneling conductance $G_T$ in series with a capacitance $C_G$
biased by a voltage source $U_{\rm ex}$. Provided the
tunneling conductance $G_T \ll  G_K$, the island charge
$q$ is quantized, i.e. $q=-ne$, where $n$ is the number of excess
electrons in the box. At zero temperature, $n$ is a staircase function of
the external voltage $U_{\rm ex}$. This idealized behavior is
modified at finite temperatures, where the occupation of higher
charge levels leads to a smearing of the staircase
\cite{lafarge}. Furthermore, the finite
tunneling conductance causes a hybridization of the lead and island
electronic states so that the island charge is no longer strictly
quantized.
Since the range of validity of perturbative results is most
restricted at zero
temperature, we focus attention here on this case.

At zero temperature, the ground state energy ${\cal E}$ as a
function of $n_{\rm ex}=C_G U_{\rm ex}/e$ determines the average
island charge
\begin{equation}
\langle n \rangle = n_{\rm ex} -
\frac{1}{2 E_c}~\frac{\partial {\cal E}}{\partial n_{\rm ex}}~~.
\label{eq:nmittel1}
\end{equation}
Here $E_c=e^2/2C$ is the classical charging energy, where the island
capacitance $C$ is the sum of the capacitance $C_T$ of the tunnel
junction and the gate capacitance $C_G$,
{\it cf}.\ Fig.~\ref{fig:fig1}. Formally ${\cal E}$ may be written as a
perturbation series in powers of the
tunneling conductance, leading to a diagrammatic
representation of ${\cal E}$ \cite{physrevB}. The
zeroth order term of the ground state energy for vanishing electron
tunneling is determined by the minimum of the electrostatic energy
and reads ${\cal E}_0=E_c(n_0-n_{\rm ex})^2$ where $n_0$ is the
integer closest to
$n_{\rm ex}$. Hence, as function of the applied voltage, the island charge
displays the well known Coulomb staircase $\langle n \rangle =
n_0$. Because of the periodicity and antisymmetry of $\langle n
\rangle$ as a function of $n_{\rm ex}$, we may
confine ourselves to $0\leq n_{\rm ex}<{\scriptstyle \frac{1}{2}}$.

\begin{figure}
\begin{center}
\leavevmode
\epsfxsize=0.35\textwidth
\epsfbox{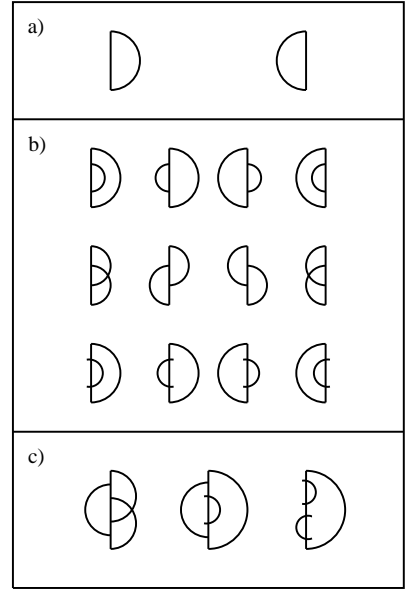}
\end{center}
\caption{Diagrams for the ground state energy in a) first order,
b) second order, and c) selected diagrams in third order in $\alpha$.}
\label{fig:fig2}
\end{figure}

Since the tunneling Hamiltonian $H_T$ transfers an electron from
the lead to
the island electrode or vice versa, it has no diagonal
components in the basis of unperturbed energy eigenstates and only
even orders in $H_T$ contribute to the perturbation series of ${\cal
E}$. The second order term gives a contribution proportional to the
dimensionless conductance $\alpha = G_T/G_K$ and is represented by  
the two
diagrams shown in Fig.~\ref{fig:fig2}a. The arc to the right
describes the formation of a virtual electron-hole pair in the
intermediate state with an excess
electron on the island, that is $n=1$, and a hole in the lead
electrode. The second diagram describes the
corresponding process with an intermediate state of reduced island
charge $n=-1$. In the interval
$0\leq n_{\rm ex} < {\scriptstyle \frac{1}{2}}$ considered,
the two diagrams give a contribution to the average island charge of
the form
\cite{matveev,esteve}
\begin{equation}
\langle n \rangle_1^{}=g\ln\frac{1+2n_{\rm
ex}}{1-2n_{\rm ex}},
\end{equation}
where $g=\alpha/4\pi^2$.
The contribution of fourth order in $H_T$ follows from the
diagrams depicted in Fig.~\ref{fig:fig2}b. Here the first eight
diagrams describe processes with two intermediate electron-hole pairs
created and annihilated in all
possible ways. The remaining diagrams have a lower order diagram
inserted, as indicated by the prolongation of the arc
across the vertical line. These diagrams correspond to terms in the
Rayleigh-Schr\"odinger perturbation series with energy denominators
squared. Each of the arcs depicts an integral over the energy of a
virtual electron-hole pair
with a spectral density which becomes
linear in the infinite bandwidth limit. An element of the vertical line
corresponds to an energy denominator describing the energy difference
between the intermediate virtual state and the ground state.
The diagrammatic rules are explained in detail in \cite{physrevB} and
it was shown there that although each
single diagram
diverges in the infinite bandwidth limit, the sum of all diagrams of a
given order remains finite. The contribution of second order in
$\alpha$ to the average island charge reads
\cite{physicaB}
\begin{equation}
\begin{array}{rl}
\langle n \rangle _2 & = -g^2\Big\{
n_{\rm ex}\left[\frac{4\pi^2}{3}+\ln^2\left(\frac{1-2n_{\rm
ex}}{1+2n_{\rm
ex}}\right)\right]    \\
    & + \frac{16(1+2n_{\rm ex}-2n_{\rm ex}^2)}{(3-2n_{\rm
ex})(1+2n_{\rm ex})} \ln(1-2n_{\rm ex})  \\
    & + 2(1-n_{\rm ex})\Big[ \ln^2\left(\frac{1-2n_{\rm
ex}}{4(1-n_{\rm ex})}\right)+ 2
\mbox{Li}_2 \left(\frac{3-2n_{\rm ex}}{4(1-n_{\rm ex})}\right)  \\
    & -\frac{8(1-n_{\rm ex})}{(1-2n_{\rm ex})(3-2n_{\rm ex})}
\ln(4(1-n_{\rm ex})) \Big] -\mbox{s.t.}(-n_{\rm ex}) \Big\}.
\end{array}
\end{equation}
Here $\mbox{s.t.}(-n_{\rm ex})$ stands for the same sum of terms with
$n_{\rm ex}$ replaced by
$-n_{\rm ex}$ showing explicitly the asymmetry of $\langle n
\rangle$ in the applied
voltage $U_{\rm ex}$. Further, $\mbox{Li}_2(x)=-\int_0^x \mbox{d}z~
\ln(1-z)/z$ denotes the dilogarithm function \cite{lewin}.

In third order in $\alpha$, one has to evaluate $160$ diagrams,
some of which are shown in Fig.~\ref{fig:fig2}c. There are $80$
diagrams without insertions, such as the left diagram in
Fig.~\ref{fig:fig2}c, $64$ diagrams with one insertion, and $16$
diagrams with two insertions. Using the diagrammatic rules, we have  
to deal with
three-fold energy integrals of  rational kernels. These integrals can
be done analytically leading to polylogarithms
\cite{lewin} and powers of logarithms of rational arguments. Since the
full analytic expression is rather involved \cite{dritte}, we present
explicitly only results
for two limiting cases $n_{\rm ex}\rightarrow 0$ and $n_{\rm
ex}\rightarrow
{\scriptstyle \frac{1}{2}}$.

For small external voltage, the average island charge grows linearly as
\begin{equation}
\langle Q \rangle = e\langle n \rangle = C^* U_{\rm ex}
\end{equation}
where $C^*$ is an effective capacitance of the box. In the absence of
Coulomb blockade effects $C^*=C_G$, while for strong Coulomb
blockade, i.e., in the limit of vanishing tunneling conductance,
$C^*=0$. It is thus natural to characterize the strength of the Coulomb
blockade effect by an effective charging energy $E_c^*$ defined by
\cite{europhysics}
\begin{equation}
\frac{E_c^*}{E_c} =1-\frac{C^*}{C}=1-
 \left. \frac{\partial \langle n \rangle }{\partial
 n_{\rm ex}} \right|_{n_{\rm ex}=0} .
\label{eq:is0}
\end{equation}
The perturbation series gives
\begin{equation}
\frac{E_c^*}{E_c} =
 1-4g+Ag^2-Bg^3+{\cal O}(g^4),
\label{eq:u}
\end{equation}
where $A=5.066...$ and $B=1.457...$ are analytically known
coefficients whose explicit form is too lengthy to present here.

\begin{figure}
\begin{center}
\leavevmode
\epsfxsize=0.35\textwidth
\epsfbox{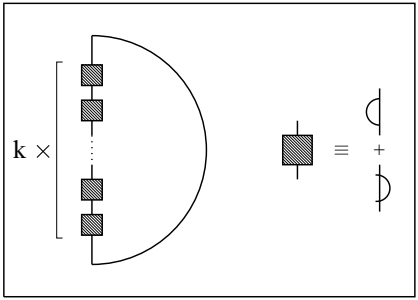}
\end{center}
\caption{Diagrams of order $\alpha^{k+1}$ for the ground state energy
giving leading order logarithms.}
\label{fig:fig3}
\end{figure}

The perturbation series is well behaved in
the region of
interest, except in the vicinity of $n_{\rm ex}={\scriptstyle
\frac{1}{2}}$, where
logarithmic divergences arise. This unphysical behavior indicates
the failure of perturbation theory due to the degeneracy of
the ground state in this limit.
For $n_{\rm ex}\rightarrow {\scriptstyle \frac{1}{2}}$ one
finds from the analytical expression for the average island charge
\begin{eqnarray}
\langle n \rangle &=&
 ag^2+bg^3-(g+6g^2+cg^3)\ln\delta     \nonumber \\
 &&-(2g^2+24g^3)\ln^2\delta
  -4g^3\ln^3\delta + {\cal O}(\delta)
\label{eq:nmittel3}
\end{eqnarray}
where $\delta = {\scriptstyle \frac{1}{2}}-n_{\rm ex}$,
and where the coefficients $a$, $b$, and $c$ read numerically
$a=-9.7726...$,
$b=-70.546...$, and
$c=65.462...$\,. The leading order logarithmic terms in
Eq.\ (\ref{eq:nmittel3}) are
$-g\ln\delta-2g^2\ln^2\delta-4g^3\ln^3\delta$. These terms come
from the diagrams shown in Fig.~\ref{fig:fig3}, where all
intermediate states are confined to the two charge
states $n=0$, $1$ which are degenerate at $n_{\rm
ex}={\scriptstyle \frac{1}{2}}$.
The most divergent logarithmic term of order $k$ reads
$-{\scriptstyle \frac{1}{2}}(2g\ln\delta)^k$, and we get by resummation  
\begin{equation}
\langle n \rangle =
\frac{-g\ln\delta}{1-2g\ln\delta}.
\label{eq:matveev}
\end{equation}
This result was previously derived by Matveev \cite{matveev} using
renormalization group techniques.

To explore the range of validity of these higher order perturbative
results we have carried out precise Quantum Monte Carlo (QMC)
simulations for the single electron box. In contrast to earlier
attempts \cite{falci,europhysics,zwerger}, we do not work in the phase
representation, but simulate configurations directly in the charge
representation, keeping track of all intermediate electron-hole
pairs. A general numeric scheme for evaluating series of integrals
directly in the continuum, i.e.,
without invoking artificial discretization of the integration
variables was explained in \cite{JETPlett96,JETP98}. It is possible
then to develop an exact (without systematic errors)
QMC algorithm summing diagrammatic series \cite{KolyaBorya}.
Here, the integration variables are imaginary times of charge
transfer events, and we have employed this ``diagrammatic" QMC for
summing all graphs
for the partition function \cite{physrevB}.
The efficiency of the new
algorithm has allowed us to simulate the single electron box at
extremely low temperatures ($\beta E_c$ as large as $10^4$). While in the
phase representation $n_{\rm ex} \neq 0$ results in a sign problem,  
in the
charge representation all contributions are positive
definite even at finite external voltage. This
has enabled us to obtain for the first time QMC data for the entire
staircase function.

\begin{figure}
\begin{center}
\leavevmode
\epsfxsize=0.45\textwidth
\epsfbox{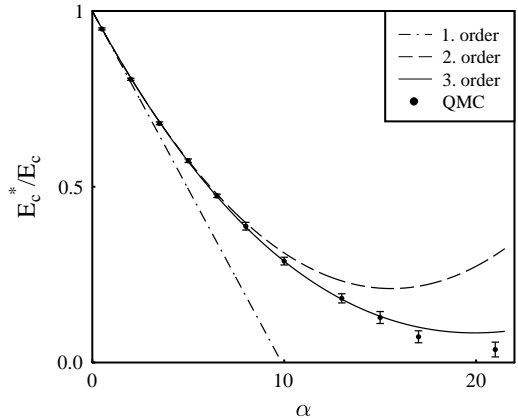}
\end{center}
\caption{Effective charging energy as a function of the dimensionless
conduction $\alpha=G_T/G_K$. Perturbative results are compared with
QMC data.}
\label{fig:fig4}
\end{figure}

In Fig.\ \ref{fig:fig4}, the effective charging energy $E_c^*$ at zero
temperature is shown as a function of the dimensionless conductance
$\alpha$. Apart from the point for $\alpha=21$, the QMC data are 
consistent with previous results by
Wang et al.\ \cite{europhysics}, obtained with a different
algorithm. For small external voltages, the analytical
result to second order in $\alpha$ is correct with errors below
$4\%$ for values of the dimensionless conductance up to
$\alpha \approx 8$, while the
third order result extends to $\alpha \approx 16$.

\begin{figure}
\begin{center}
\leavevmode
\epsfxsize=0.45\textwidth
\epsfbox{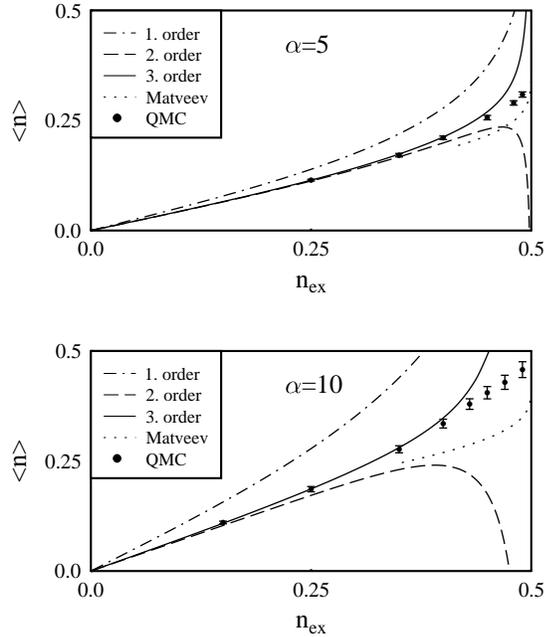}
\end{center}
\caption{The average electron number $\langle n \rangle$ as a
function of the dimensionless voltage $n_{ex}$ is shown in first, second,
and third order perturbation theory in $\alpha$, and compared with QMC
data. The result (8) is also shown as a dotted line.}
\label{fig:fig5}
\end{figure}

The range of validity of perturbation theory shrinks with increasing
external voltage and is worst near $n_{\rm ex}={\scriptstyle
\frac{1}{2}}$. In Fig.\ \ref{fig:fig5},
the rounded Coulomb staircase is depicted for $\alpha = 5$ and $\alpha
= 10$. While the analytical result diverges at
$n_{\rm ex} = {\scriptstyle \frac{1}{2}}$, we
find that third order perturbation theory in
$\alpha$ remains valid with errors
below $4\%$ up to $n_{\rm ex} \approx 0.495$ for dimensionless conductance
$\alpha=2$ (data
not shown), up to $n_{\rm ex} \approx 0.45$ for $\alpha=5$, and
up to $n_{\rm ex} \approx 0.4$ for $\alpha=10$. Since for $n_{\rm ex}=0.45$
the charging energies for $n=0$ and
$n=1$ differ only by $0.1 E_c$, deviations from the third order result
in $\alpha$ can be observed only for temperatures well below
$E_c/10k_B$ even at $n_{\rm ex}={\scriptstyle \frac{1}{2}}$. Finally,
Fig.\ \ref{fig:fig5} shows that the resummation of the
leading
logarithmic terms (\ref{eq:matveev}) does not suffice to describe
the behavior
near $n_{\rm ex} = {\scriptstyle \frac{1}{2}}$. Subleading logarithms
are important to obtain quantitatively meaningful results.

In summary, we have studied the breakdown of Coulomb charging effects
with increasing tunnel conductance. Precise Monte Carlo data for
the effective charging energy as well as for the smeared staircase
function were presented. Comparing analytical results with
these data, the range of validity of expansions in
powers of the tunneling conductance was determined. It was found that
presently attainable experimental results are covered by expansions  
up to third
order in $\alpha$. With increasing $\alpha$, the effective
charging energy $E_c^*$ decreases rapidly. Since finite temperature
corrections for large $\alpha$ are controlled by the dimensionless
parameter $\beta E_c^*$\cite{europhysics}, experiments with
dimensionless conductance above $\alpha=10$, where perturbation theory
begins to fail, require extremely low temperatures to see nonperturbative
effects. We finally mention that the methods employed here can be
extended to other systems displaying charging
effects, such as the single electron transistor.

The authors would like to thank M.H.\ Devoret, D.\ Esteve, P.\
Joyez, J.\ K\"onig, and H.\ Schoeller for valuable discussions.  
H.G.\ and N.P.\ acknowledge hospitality and support by the Institute
for Nuclear Theory, Washington University, Seattle, where this work
was started. Additional support was provided by the Deutsche
Forschungsgemeinschaft (Bonn) and the Russian Foundation for Basic  
Research.

\end{multicols}
\end{document}